\documentclass[10pt,a4paper]{article}
\usepackage{latexsym,epsfig,multicol}
\usepackage{color}
\usepackage{graphicx}
\usepackage{pifont}
\usepackage{verbatim}

\def\beq{\begin{eqnarray}}
\def\eq{\end{eqnarray}}

\makeatletter

\newenvironment{figurehere}
  {\def\@captype{figure}}
  {} 
\makeatother

\usepackage{amssymb}
\usepackage{amsfonts}
\usepackage{amsmath}%
\usepackage{graphicx}

\providecommand{\U}[1]{\protect\rule{.1in}{.1in}}

\title{ Spontaneous parity violation and minimal Higgs models.}
\author{H. Chavez and J. A. Martins Sim\~{o}es,\\Instituto de F\'{\i}sica,\\Universidade Federal do Rio de Janeiro, RJ, Brazil}
\date{}

\begin{document}

\maketitle

\begin{abstract}
In this paper we present a model for the spontaneous breaking of parity with
two Higgs doublets and two neutral Higgs singlets which are even and odd under
$\mathcal{D}$-parity. The condition $\upsilon_{R}\gg\upsilon_{L}$ can be
satisfied without introducing bidoublets and it is induced by the breaking of
$\mathcal{D}$-parity through the vacuum expectation value of the odd Higgs singlet.
 Examples of left-right symmetric  and  mirror fermions models in grand unified 
theories are presented.


\end{abstract}

\section{Introduction}

Left-right symmetric models with spontaneous parity breaking offer a natural
explanation for the parity asymmetry observed in nature. The gauge group
$SU(2)_{L}\otimes SU(2)_{R}\otimes U(1)_{B-L}$ fixes the interactions and
generalizes the standard electroweak theory. However the fundamental fermionic
representation and the Higgs sector are not completely determined. For the
fermions one can have two possibilities: new right-handed doublets, as in the
earlier models\cite{MOA}, or new mirror fermions\cite{MAA,NOS}. The Higgs
sector has more possibilities and introduces more unknown parameters in the
model. It is highly desirable to have the minimum number of unknown parameters
in order to compare models and experimental data. A recent work in this
direction was done by Brahmachari, Ma and Sarkar \cite{BRA}. With two Higgs
doublets, ${\chi}_{R}$ and $\chi_{L}$, and a dimension five operator they have
proposed a left-right model including fermion masses. Another mirror model
with two Higgs doublets and two Higgs singlets was developed in ref.3. In both
cases the condition
\begin{equation}
\upsilon{_{R}\gg\upsilon_{L}}%
\end{equation}
must be satisfied. The present experimental bound is $\upsilon_{R}%
>30\,\upsilon_{L}$ \cite{NOS}. Later on, Siringo \cite{SIR} revived an earlier
remark by Senjanovic and Mohapatra \cite{SEN} that the above condition can not
be satisfied in models with only two Higgs doublets. These remarks seems to
leave open the possibility that only scalar bidoublets could break parity in a
consistent way. This possibility exists but has the unpleasant feature of a
large number of Higgs fields and undetermined parameters.
\par
However, there is other elegant way \cite{Chang} to produce the condition (1)  by
introducing one singlet Higgs which is odd under $\mathcal{D}$-parity. The
difference with the $\mathcal{P}$-parity breaking is that in $\mathcal{D}$-parity the
 vacuum expectation value (VEV) of a
parity odd field can  be spontaneously broken without breaking
the left-right symmetry. In consequence, the gauge coupling constants of
$SU(2)_{R}$ and $SU(2)_{L}$ also can be different and  left-handed
and right-handed scalars can have different masses and VEVs.
\par
In the present paper, we extend the previous analysis to include also a
singlet Higgs field which is even under $\mathcal{D}$-parity and mixes
with the odd field. We show that this mixing term also contribute to the
hierarchy relation (1). We include in our study two examples of grand unified theories 
where Higgs singlets transforming under $\mathcal{D}$-parity  are assigned  to
the L-R symmetric model and to a mirror fermion model.

\section{The scalar potential and the breaking of the L-R symmetry}

There are two forms of breaking parity spontaneously: the first is to identify the
discrete symmetry $Z_{2}$ that interchanges the groups $SU(2)_{L}$ \ and
$SU(2)_{R}$ of the Lorentz group $O(3,1)$ as the parity operator $\mathcal{P}$
that transforms the Higgs bosons $\chi_{L}\ \ \underleftrightarrow
{\mathcal{P}}\ \chi_{R}$ and also $W_{L}\ \ \underleftrightarrow{\mathcal{P}%
}\ W_{R}$ . So, when $SU(2)_{R}$ is broken in the symmetric L-R model, parity $\mathcal{P}$ is also broken.
 The second possibility of spontaneously breaking the
parity symmetry is through  the VEV of an odd scalar field which preserves
L-R symmetry. This type of parity is called $\mathcal{D}$-parity
which is a generator of larger groups that contain the product $SU(2)_{L}\otimes
SU(2)_{R}$ as a subgroup. This second possibility is very interesting because it 
allows $\left\langle \chi_{L}\right\rangle \ll\left\langle \chi_{R}%
\right\rangle $ with different coupling constants for $SU(2)_{L}$ and
$SU(2)_{R}$ and different masses for the Higgs fields.
\par

Our model for the scalar potential includes two doublets and two singlets Higgs fields.
These singlets and doublets transforms  under $\mathcal{D}$-parity as
$S_{M}\ \underleftrightarrow{\mathcal{D}}\ S_{M}$ ; $S_{D}%
\ \underleftrightarrow{\mathcal{D}}-S_{D}$ and $\chi_{L}\ \underleftrightarrow
{\mathcal{D}}\ \chi_{R}$ , if in the model there are no CP violating terms  or no 
complex Yukawa couplings. We propose the following invariant
potential under $G_{3221}=SU(3)_{C}\otimes SU(2)_{L}\otimes SU(2)_{R}\otimes
U(1)_{B-L}$ for the Higgs fields%

\begin{gather}
V(\chi_{L},\chi_{R},S_{D},S_{M})=\mu^{2}({\chi_{L}^{\dagger}}\chi_{L}%
+{\chi_{R}^{\dagger}}\chi_{R})-\lambda_{\chi}({\chi_{L}^{\dagger}}\chi
_{L}+{\chi_{R}^{\dagger}}\chi_{R})^{2}-m_{D}^{2}S_{D}^{2}-\nonumber\\
\eta_{D}S_{D}^{3}-\lambda_{D}S_{D}^{4}-m_{M}^{2}S_{M}^{2}-\eta_{M}S_{M}%
^{3}-\lambda_{M}S_{M}^{4}+M_{D}S_{D}({\chi_{R}^{\dagger}}\chi_{R}-{\chi
_{L}^{\dagger}}\chi_{L})+\nonumber\\
M_{M}S_{M}({\chi_{L}^{\dagger}}\chi_{L}+{\chi_{R}^{\dagger}}\chi_{R})+\lambda
S_{D}S_{M}({\chi_{R}^{\dagger}}\chi_{R}-{\chi_{L}^{\dagger}}\chi
_{L})+\nonumber\\
(\varepsilon_{D}S_{D}^{2}+\varepsilon_{M}S_{M}^{2})({\chi_{L}^{\dagger}}%
\chi_{L}+{\chi_{R}^{\dagger}}\chi_{R})-\varkappa(({\chi_{L}^{4})}^{\dagger}%
{+}\chi_{L}^{4}+({\chi_{R}^{4})}^{\dagger}{+}\chi_{R}^{4}).
\end{gather}
Our motivation for  this potential is the fact that $S_{M}$ and
$S_{D}$ do not necessarily belong to the same irreducible multiplet of Higgs
fields. In consequence it is also possible that these fields are mixed. If this is
the case, when $\left\langle S_{D}\right\rangle =s_{D}$\ and $\left\langle
S_{M}\right\rangle =s_{M}$ the  potential terms that contributes to the masses of Higgs
fields $\chi_{L}$ and $\chi_{R}$\ are

\begin{align}
V_{\text{\textbf{mass}}}(\chi_{L},\chi_{R})  &  =(\mu^{2}+\varepsilon_{D}%
s_{D}^{2}+\varepsilon_{M}s_{M}^{2}+M_{M}s_{M})(\left\vert \chi_{L}\right\vert
^{2}+\left\vert \chi_{R}\right\vert ^{2})+\nonumber\\
&  (M_{D}s_{D}+\lambda s_{D}s_{M})(\left\vert \chi_{R}\right\vert
^{2}-\left\vert \chi_{L}\right\vert ^{2}),
\end{align}
from which we obtain the masses
\begin{align}
m_{R}^{2}  &  =\mu^{2}+\varepsilon_{D}s_{D}^{2}+\varepsilon_{M}s_{M}^{2}%
+M_{M}s_{M}+M_{D}s_{D}+\lambda s_{D}s_{M},\\
m_{L}^{2}  &  =\mu^{2}+\varepsilon_{D}s_{D}^{2}+\varepsilon_{M}s_{M}^{2}%
+M_{M}s_{M}-M_{D}s_{D}-\lambda s_{D}s_{M}.
\end{align}
Now we impose the hierarchy condition in the previous equations such that
$m_{R}^{2}$ $\ll s_{D}^{2}\ll s_{M}^{2}$.It is necessary to indicate that $\upsilon_{L}$ breaks
the electroweak symmetry and $\upsilon_{R}$ breaks the L-R symmetry close to
the TeV scale.  So we can have, for example
$\left\langle \chi_{L}\right\rangle =\upsilon_{L}\sim m_{L}\sim100GeV$ and,
; $\ \left\langle \chi_{R}\right\rangle =\upsilon_{R}\sim m_{R}%
\sim10TeV\gg\upsilon_{L}.$ It also must be noted that if $S_{D}$\ and $S_{M}$ are in the
same multiplet, then several possible mixing terms in the potential possibility are absent.
\par

Let us now suppose that there are no CP violating terms and that all VEVs are considered to be real. With $\left\langle \chi_{L}\right\rangle =\dbinom
{0}{\upsilon_{L}},$ $\left\langle \chi_{R}\right\rangle =\dbinom{0}%
{\upsilon_{R}},$ it is possible to show that the minimum conditions for the
potential are given by
\begin{gather}
\frac{\partial V}{\partial\upsilon_{L}}=2\upsilon_{L}[\mu^{2}-2\lambda_{\chi
}(\upsilon_{L}^{2}+\upsilon_{R}^{2})-M_{D}s_{D}+M_{M}s_{M}-\lambda s_{D}%
s_{M}+\nonumber\\
\varepsilon_{D}s_{D}^{2}+\varepsilon_{M}s_{M}^{2}-4\varkappa\upsilon_{L}%
^{2}]=0,\ \\
\frac{\partial V}{\partial\upsilon_{R}}=2\upsilon_{R}[\mu^{2}-2\lambda_{\chi
}(\upsilon_{L}^{2}+\upsilon_{R}^{2})+M_{D}s_{D}+M_{M}s_{M}+\lambda s_{D}%
s_{M}+\nonumber\\
\varepsilon_{D}s_{D}^{2}+\varepsilon_{M}s_{M}^{2}-4\varkappa\upsilon_{R}%
^{2}]=0,
\end{gather}
From these equations we have
\begin{equation}
\upsilon_{L}\frac{\partial V}{\partial\upsilon_{R}}-\upsilon_{R}\frac{\partial
V}{\partial\upsilon_{L}}=4\upsilon_{L}\upsilon_{R}[M_{D}s_{D}+\lambda
s_{D}s_{M}-2\varkappa(\upsilon_{R}^{2}-\upsilon_{L}^{2})]=0
\end{equation}
Now we require non trivial solutions such that $\upsilon_{L}\neq$ $\upsilon_{R}%
\neq0.$ Thus we obtain the desired hierarchy 
\begin{equation}
\upsilon_{R}^{2}-\upsilon_{L}^{2}=\frac{s_{D}(M_{D}+\lambda s_{M})}%
{2\varkappa}.
\end{equation}
A important point to be noted in the previous equation is that the breaking effect  due to the singlet $S_{M}$ is sub-dominant with relation to $S_{D}$
that breaks  $D$-parity when $\left\langle S_{D}\right\rangle =s_{D}$ .
Additionally, if $ s_{D}=0$ then $D$-parity is conserved and the L-R
symmetry condition is recovered, $\upsilon_{R}=\upsilon_{L}$.
We have showed that the in our potential it is possible to 
construct models with L-R symmetry and producing an hierarchy
between the breaking scale of $SU(2)_{R}$ and the electroweak scale. The main ingredient is the presence of two
Higgs singlets to generate the minimum of the potential. The crucial point in
this sense is the inclusion of the mixing term $\lambda S_{D}S_{M}({\chi
_{R}^{\dagger}\chi}_{R}{-}\chi_{L}^{\dagger}\chi_{L})$ which is possible if
$\ S_{M}$ and $S_{D}$ belong to different irreducible representations. As in the previous term, also the term $M_{D}S_{D}({\chi_{R}^{\dagger}}\chi_{R}-{\chi_{L}^{\dagger}}\chi_{L})$ breaks the L-R symmetry . It is also fundamental the fine tuning of the parameters of the model at the tree
level in order to assure that $\upsilon_{R}$ do not destabilizes the $\upsilon_{L}$ value.
Thus, from  equations (5) - (7) we have
\begin{equation}
m_{L}^{2}-2(\lambda_{\chi}+2\varkappa)\upsilon_{L}^{2}=2\lambda_{\chi}%
\upsilon_{R}^{2}\,
\end{equation}

\section{An $SO(10)$  L-R symmetric model.}

There is a known GUT context to embed the L-R symmetric model which is based on
$SO(10)$ trough its maximal sub-group as done by the Pati-Salam \cite{MOA}
approach $G_{PS}=SU(4)_{C}\otimes SU(2)_{L}\otimes SU(2)_{R}$. The idea
consists to break $\mathcal{D}$-parity below the breaking of $SO(10)$
as showed in the following breaking chain 

\begin{gather}
SO(10)\quad\underrightarrow{S_{M}}\quad G_{PS}\otimes\mathcal{D}%
\quad\underrightarrow{S_{D}}\quad SU(3)_{C}\otimes SU(2)_{L}\otimes
SU(2)_{R}\otimes U(1)_{B-L}\quad\nonumber\\
\underrightarrow{\chi_{R}}\quad SU(3)_{C}\otimes SU(2)_{L}\otimes
U(1)_{Y}\quad\underrightarrow{\chi_{L}}\quad SU(3)_{C}\otimes U(1)_{e.m},
\end{gather}
The quantum numbers for the Higgs representations that produces the pattern
(11) are given in the Table 1.
\begin{gather*}
\text{\textbf{Table 1 \ }}\\%
\begin{tabular}
[c]{|c|c|c|c|}\hline
$S_{M}\sim$ & $\{\mathbf{54\}}$ & $\supset\lbrack\mathbf{1,1,1]}$ &
$\sim\mathbf{(1,1,1,}0),$\\\hline
$S_{D}\sim$ & $\mathbf{\{45\}}$ & $\mathbf{\supset\lbrack15,1,1]}$ &
$\mathbf{\supset(1,1,1},0\mathbf{),}$\\\hline
$\chi_{R}\sim$ & $\mathbf{\{144}^{\ast}\mathbf{\}}$ & $\mathbf{\supset
\lbrack4,1,2]}$ & $\mathbf{\supset(1},\mathbf{1},\mathbf{2},-1\mathbf{)}
,$\\\hline
$\chi_{L}\sim$ & $\mathbf{\{144\}}$ & $\mathbf{\supset\lbrack4,2,1]}$ &
$\mathbf{\supset(1,2,1,-}1),$\\\hline
\end{tabular}
\\
\text{Higgs representations for the  breaking chain (11).}
\end{gather*}
Our notation is as follows: the representations between $\mathbf{\{\ \}}$
corresponds to $SO(10),$ $[\ ]$ corresponds to $G_{PS}$ and those with $(\ )$ \ corresponds
to $G_{3221}=SU(3)_{C}\otimes SU(2)_{L}\otimes SU(2)_{R}\otimes U(1)_{B-L}$
$.$ In our model a different point  from the approach of  Ref.\cite{1} is that we
are using the singlet component $[\mathbf{1,1,1]}$ of $\{\mathbf{54\}}$ in order to
break $SO(10)$ down to $G_{PS}\otimes\mathcal{D}$ because this is a symmetric
representation $\mathcal{D}$-even \cite{Chang}\cite{Mohapatra}$,$ different
of\ the $[\mathbf{1,1,1]}$ component of $\mathbf{\{210\}}$ which is
$\mathcal{D}$-odd, as required of our analysis of the previous section. Note
also that the neutral component $\mathbf{(1,1,1},0\mathbf{)\subset
\lbrack15,1,1]\subset\{45\}}$ of $\ SO(10)$ is $\mathcal{D}$-odd under $G_{3221}.$
Thus, we are expecting that the VEV of $S_{D}\sim\mathbf{\{45\},}$ which will
induce the breaking  of the L-R symmetry, depend of the VEV of $\ S_{M}.$ This
choice could allow the breaking of the L-R symmetry   close to the
electroweak scale, let us say in the few TeV scale.

The $G_{3221}$ invariant Higgs potential of  Equation.(2)  could come from the
following $SO(10)$ potential
\begin{gather}
\mathcal{L}=\mu^{2}(\mathbf{144}^{\ast}\times\mathbf{144})+\lambda_{\chi
}(\mathbf{144}^{\ast}\times\mathbf{144})^{2}+m_{D}^{2}(\mathbf{45}%
)^{2}+\nonumber\\
\eta_{D}(\mathbf{45})^{3}+\lambda_{D}(\mathbf{45})^{4}+m_{M}^{2}%
(\mathbf{54})^{2}+\eta_{M}(\mathbf{54})^{3}+\lambda_{M}(\mathbf{54})^{4}\nonumber\\
+M_{D}(\mathbf{45})(\mathbf{144}^{\ast}\times\mathbf{144})+
M_{M}(\mathbf{54})(\mathbf{144}^{\ast}\times\mathbf{144})\nonumber\\+\lambda
(\mathbf{54}\times\mathbf{45})(\mathbf{144}^{\ast}\times\mathbf{144}
)+
(\varepsilon_{D}(\mathbf{45})^{2}\nonumber\\+\varepsilon_{M}(\mathbf{54})^{2}
)(\mathbf{144}^{\ast}\times\mathbf{144})+\varkappa\lbrack(\mathbf{144}^{\ast
})^{4}+(\mathbf{144})^{4}].
\end{gather}

Let us notice that the term $(\mathbf{54}\times\mathbf{45})(\mathbf{144}^{\ast
}\times\mathbf{144})$ is possible if the interactions between $(\mathbf{144}^{\ast
}\times\mathbf{144})$ and $(\mathbf{54}\times\mathbf{45})$ are mediated by
the gauge boson in the $\{\mathbf{45\}}$ or $\{\mathbf{54\}}$ representations.

The corresponding hypercharges are given by
\begin{equation}
\frac{Y}{2}=T_{3R}+\frac{B-L}{2}.
\end{equation}
The left-handed ordinary fermions are contained in

\begin{gather}
\mathbf{\{16\}}_{iL}\mathbf{=}%
\genfrac{}{}{0pt}{0}{\mathbf{[4,2,1]}}{\overbrace{q_{L}=\left(
\genfrac{}{}{0pt}{0}{u}{d}\right)  _{L}(\mathbf{3,2,1,\,}1/3\mathbf{)}\oplus
l_{L}=\dbinom{\nu}{e}_{L}(\mathbf{1,2,1,}-1\mathbf{\,)}}}%
\oplus\nonumber\\%
\genfrac{}{}{0pt}{0}{\mathbf{[4}^{\ast}\mathbf{,1,2]}}{\overbrace{q^{C}%
\,_{L}\,=\dbinom{d^{C}}{u^{C}}_{L}(\mathbf{3}^{\ast}\mathbf{,1,2,\,}%
-1/3\mathbf{)}\oplus l^{C}\,_{L}\,=\dbinom{e^{C}}{\nu^{C}}_{L}%
(\mathbf{1,1,2,\,}1\mathbf{)}}}%
.
\end{gather}
\qquad\qquad

The Majorana and Dirac masses in our model can  arise from dimension-5
effective operators such as%

\begin{gather*}
\mathcal{O}_{1}=\frac{1}{\Lambda_{Q}}\left(  \overline{q_{L}}\chi_{L}\right)
\left(  \overline{q_{R}}\chi_{R}^{\ast}\right)  ,\qquad\mathcal{O}_{2}%
=\frac{1}{\Lambda_{Q}}\left(  \overline{q_{L}}\chi_{L}^{\ast}\right)  \left(
q_{R}\chi_{R}\right)  ,\\
\mathcal{O}_{3}=\frac{1}{\Lambda_{D}}\left(  \overline{l_{L}}\chi_{L}\right)
\left(  l_{R}\chi_{R}^{\ast}\right)  ,\qquad\mathcal{O}_{4}=\frac{1}%
{\Lambda_{D}}\left(  \overline{l_{L}}\chi_{L}^{\ast}\right)  \left(  l_{R}%
\chi_{R}\right)  ,\\
\mathcal{O}_{5}=\frac{1}{\Lambda_{M}}\left(  l_{L}\chi_{L}^{\ast}\right)
\left(  l_{L}\chi_{L}^{\ast}\right)  ,\qquad\mathcal{O}_{6}=\frac{1}%
{\Lambda_{M}}\left(  l_{R}\chi_{R}^{\ast}\right)  \left(  l_{R}\chi_{R}^{\ast
}\right)  ,
\end{gather*}
where $\Lambda_{Q,D,M}$ are the masses in the GUT scale. The first two operators
$\mathcal{O}_{1}$ and $\mathcal{O}_{2}$ give the quark masses from which it is
assumed the existence of the matter fields $\mathcal{Q}_{L,R}\sim(\mathbf{3,1,1,}%
4/3)\subset(\mathbf{15,1,1)}\subset\{\mathbf{45}\}$ or $\{\mathbf{120}\}$ in order 
to generate the operator $\mathcal{O}_{1}$ and $\mathcal{M}_{L,R}%
\sim(\mathbf{3,1,1,}-2/3)\subset(\mathbf{6,1,1)}\subset\{\mathbf{10}\}$,
$\{\mathbf{126}\}$ or $\mathcal{M}_{L,R}\sim(\mathbf{3,1,1,}-2/3)\subset
(\mathbf{10,1,1)}\subset\{\mathbf{120}\}$ in order to  generate the operator
$\mathcal{O}_{2}$.\ On the other side, the operators $\mathcal{O}_{3}$ and
$\mathcal{O}_{4}$ can be generated by the fermionic matter fields
$\mathcal{P}_{L,R}\sim(\mathbf{1,1,1,}0)\subset(\mathbf{1,1,1)}\subset
\{\mathbf{54}\}$ and $\mathcal{S}_{L,R}\sim(\mathbf{1,1,1,}-2)\subset
(\overline{\mathbf{10}}\mathbf{,1,1)}\subset\{\mathbf{120}\}.$ This is showed
in Figure 1.
\begin{figurehere}
\begin{center}
\includegraphics[width=10cm,height=5cm]{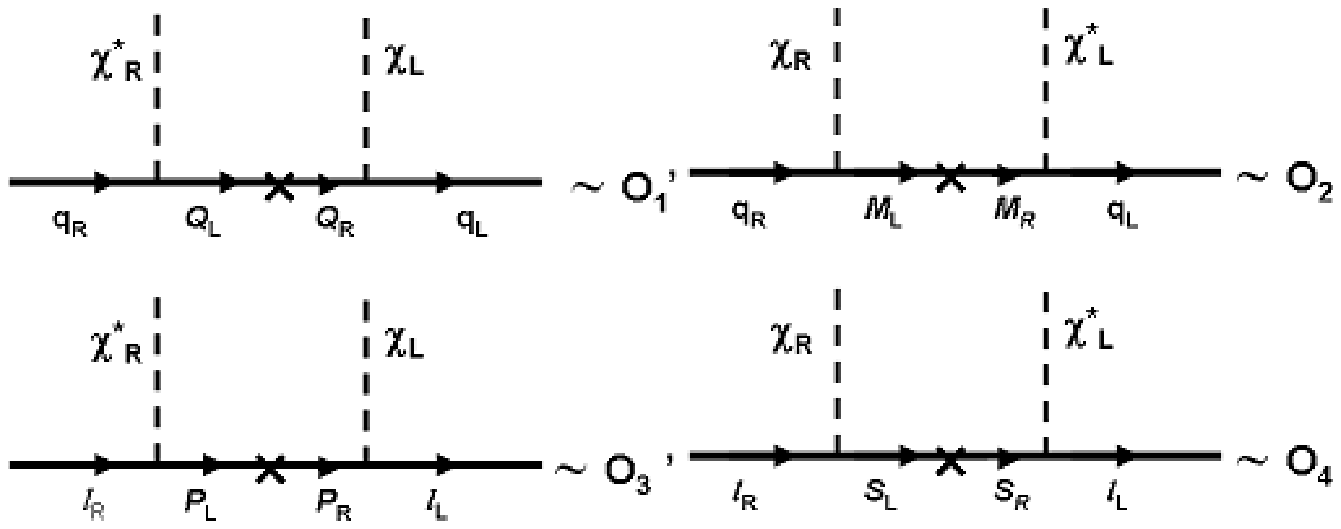}
\caption{Diagrams producing the $ \mathcal{O}_{1-4}$ operators of 
dimension-5. }
\end{center}
\end{figurehere}

\par

To obtain the operators $\mathcal{O}_{5}$ and $\mathcal{O}_{6}$ let us
observe that the operator \\ 
$(\{\mathbf{16}\}_{L}\{\mathbf{16}\}_{L}
)(\{\mathbf{144}^{\ast}\}\{\mathbf{144}^{\ast}\})$ can  be obtained through
 the mediation of the fermions in the $\{\mathbf{45}\}$ and $\{\mathbf{210}%
\}$ representations as it is showed in Figure 2. Operators of dimension-5 of the
 Majorana type$\mathcal{O}_{5}$ can be obtained from  diagram (2a), where it is
necessary the inclusion of the fermionic matter term $\{\mathbf{45}\}$ in the
$\mathbf{(15,1,1)}$ component. This contribution to the neutrino mass is large
because this fermion term corresponds to the GUT scale. To implement the see-saw
mechanism  we need to consider a term showed in Figure 2b which include the D-parity effect
through  the Higgs singlet  $\{\mathbf{45}\}$ and also a term that preserves
it through the Higgs singlet $\{\mathbf{54}\}.$ Details of the calculation
and their relevance  to generate magnetic moments for  neutrinos and charged leptons
 will be presented elsewhere\cite{Chavez}.

\begin{figurehere}
\begin{center}
\includegraphics[width=10cm,height=4cm]{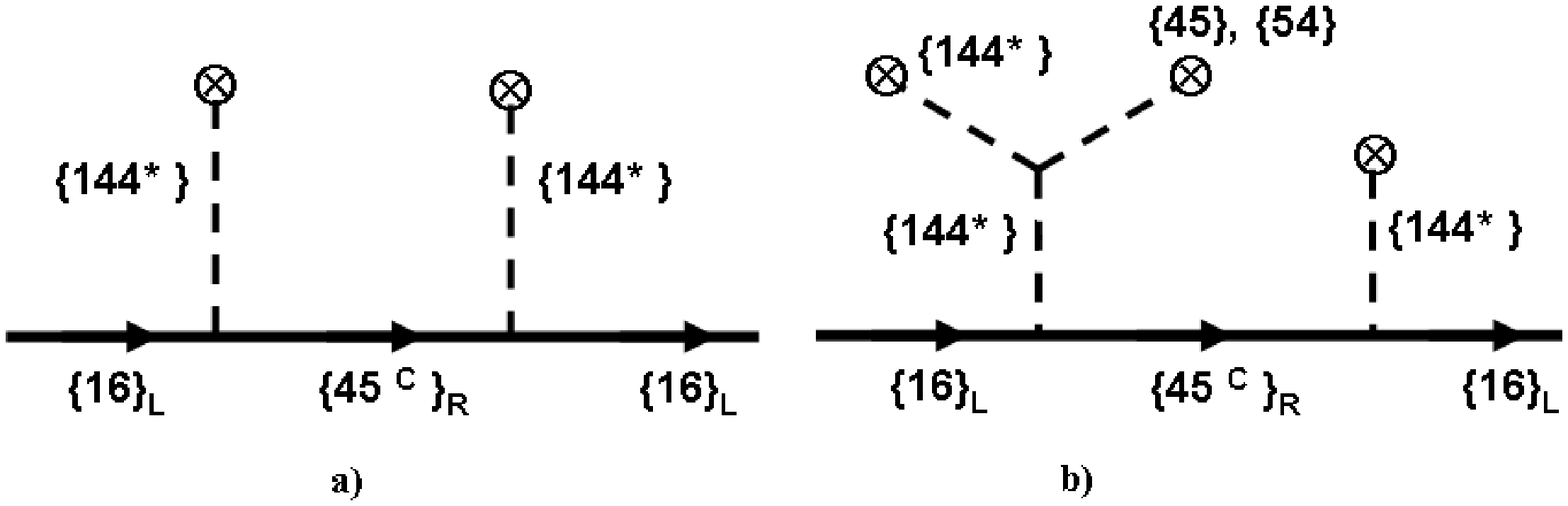}
\caption{Diagrams for  generating neutrino masses.}
\end{center}
\end{figurehere}

\par
We have another possibility to embed  $SU(2)_{L}\otimes SU(2)_{R}\otimes U(1)_{B-L}$ based in the breaking chain \cite{kau}
\begin{gather}
SO(10)\quad\underrightarrow{S_{M}}\quad G_{PS}\quad\underrightarrow{S_{D}%
}\quad SU(3)_{C}\otimes SU(2)_{L}\otimes SU(2)_{R}\otimes U(1)_{B-L}%
\quad\nonumber\\
\underrightarrow{\chi_{R}}\quad SU(3)_{C}\otimes SU(2)_{L}\otimes
U(1)_{Y}\quad\underrightarrow{\chi_{L}}\quad SU(3)_{C}\otimes U(1)_{e.m},
\end{gather}
The Higgs fields are given in table 2. The component
$\mathbf{[1,1,1]}$ of $\mathbf{\{210\}}$ is $\mathcal{D}$-odd. 
\begin{gather*}
\text{\textbf{Table 2 \ }}\\%
\begin{tabular}
[c]{|c|c|c|c|}\hline
$S_{M}\sim$ & $\mathbf{\{210\}}$ & $\supset\lbrack\mathbf{1,1,1]}$ &
$\sim\mathbf{(1,1,1,}0),$\\\hline
$S_{D}\sim$ & $\mathbf{\{210\}}$ & $\mathbf{\supset\lbrack15,1,1]}$ &
$\mathbf{\supset(1,1,1},0\mathbf{),}$\\\hline
$\chi_{R}\sim$ & $\mathbf{\{16}^{\ast}\mathbf{\}}$ & $\mathbf{\supset
\lbrack4,1,2]}$ & $\mathbf{\supset(1},\mathbf{1},\mathbf{2},-1\mathbf{)}%
,$\\\hline
$\chi_{L}\sim$ & $\mathbf{\{16\}}$ & $\mathbf{\supset\lbrack4,2,1]}$ &
$\mathbf{\supset(1,2,1,-}1),$\\\hline
\end{tabular}
\\
\text{Higgs representations for the  breaking chain (15).}%
\end{gather*}%
If we use only the fields of $\mathbf{\{210\}}$ in order to produce the first two  steps in
(15), then the  potential analogous to (12) has to be modified to%
\begin{gather}
\mathcal{L}=\mu^{2}(\mathbf{16}^{\ast}\times\mathbf{16})+\lambda_{\chi
}(\mathbf{16}^{\ast}\times\mathbf{16})^{2}\nonumber\\
+m_{M}^{2}(\mathbf{210})^{2}+\eta_{M}(\mathbf{210})^{3}+\lambda_{M}%
(\mathbf{210})^{4}+M_{M}(\mathbf{210})(\mathbf{16}^{\ast}\times\mathbf{16}%
)+\nonumber\\
\varepsilon_{M}(\mathbf{210})^{2}\times(\mathbf{16}^{\ast}\times
\mathbf{16})+\varkappa\lbrack(\mathbf{16}^{\ast})^{4}+(\mathbf{16})^{4}].
\end{gather}
Then, following ref.\cite {kau}, the masses of the Higgs doublets  obtained
from (16) are given by
\begin{align}
m_{R}^{2}  &  =\mu^{2}+M_{M}s_{M}+\varepsilon_{M}s_{M}^{2},\nonumber\\
& \\
m_{L}^{2}  &  =\mu^{2}-M_{M}s_{M}+\varepsilon_{M}s_{M}^{2},\nonumber
\end{align}
As we have  $\left\langle \chi_{L}\right\rangle =\upsilon_{L}\sim m_{L}$ and
$\left\langle \chi_{R}\right\rangle =\upsilon_{R}\sim m_{R}$ we find, after
the tunning of the model parameters, that $\upsilon_{L}\sim100GeV$ and
$\upsilon_{R}\sim GUT$ . In fact,we have the relation%
\begin{equation}
\upsilon_{R}^{2}-\upsilon_{L}^{2}=\frac{M_{M}s_{M}}{2\varkappa},
\end{equation}
From this equation we see that, due to the breaking of $\mathcal{D}$-parity in
the GUT scale by the field $\mathcal{D}$-odd $\mathbf{[1,1,1]\subset\{210\},}$
the breaking of the L-R symmetry is also induced  close to the GUT scale. This is
a different prediction of the  model given by the breaking chain (11) in which it
is possible that the breaking  of L-R symmetry occurs close to the $TeV$ scale.
Other differences are also possible from the renormalization groups equations
(RGE). This analysis for the breaking chain (15) was done in Ref.9 at one and
two-loops on the gauge couplings. Some stages of the breaking chain (11) was analyzed in
different papers \cite{Amelino-Camelia} in the same context of the RGE. In fact,
the use of the $G_{SM}$ singlets in the $\mathbf{\{16\},\{126\}}$ and
$\mathbf{\{144\}}$ representations would lead to the result that the
unification of the $G_{SM}$ coupling constants is inconsistent with the
low-energy data on these couplings. On the other side, the use of
$\mathbf{\{54\}}$ in the breaking of $SO(10)\quad\underrightarrow{S_{M}}\quad
G_{PS}\otimes\mathcal{D}$ seems to be consistent with the experimental bound
on the proton lifetime only at a marginal level \cite{Amelino-Camelia} . We 
conclude that it would be necessary a more complete analysis of the RGE for the
breaking chains along the lines of Ref. \cite{Csikor}.

\section{An L-R model based in SU(7) with mirror fermions.}

In the L-R model with mirror fermions the particle content is described in
Table 3 for the two first families with its
its quantum\ numbers under $ SU(3)_{C}\otimes SU(2)_{L}\otimes
SU(2)_{R}\otimes U(1)_{Y}$.
\begin{gather*}
\text{\textbf{Table 3}}\\%
\begin{tabular}
[c]{|c|c|}\hline
\textbf{Ordinary fermions} & \textbf{Mirror fermions }\\\hline
$%
\begin{array}
[c]{c}%
l_{L}=\dbinom{\nu_{e}}{e}_{L},\dbinom{\nu_{\mu}}{\mu}_{L}\sim(\mathbf{1,2,1,}%
-1\mathbf{\,)}\\
\mathbf{\ }e_{R},\mu_{R}\sim(\mathbf{1,1,1},-2)\\
\mathbf{\ }\nu_{eR},\nu_{\mu R}\sim(\mathbf{1,1,1},0)\\
\dbinom{u}{d}_{L},\dbinom{c}{s}_{L}\sim(\mathbf{3,2,1,}1/3\mathbf{\,)}\\
u_{R},c_{R}\sim(\mathbf{3,1,1,}4/3\mathbf{\,)}\\
d_{R},s_{R}\sim(\mathbf{3,1,1,}-2/3\mathbf{\,)}%
\end{array}
$ & $%
\begin{array}
[c]{c}%
L_{R}=\dbinom{N_{E}}{E}_{R},\dbinom{N_{M}}{M}_{R}\sim(\mathbf{1,1,2,}%
-1\mathbf{\,)}\\
\mathbf{\ }E\,_{L},\mathbf{\ }M_{L}\sim(\mathbf{1,1,1},-2)\\
N_{EL},N_{ML}\sim(\mathbf{1,1,1},0)\\
\dbinom{U}{D}_{R},\dbinom{C}{S}_{R}\sim(\mathbf{3,1,2,}1/3\mathbf{\,)}\\
U_{L},C_{L}\sim(\mathbf{3,1,1,}4/3\mathbf{\,)}\\
D_{L},S_{L}\sim(\mathbf{3,1,1,}-2/3\mathbf{\,)}%
\end{array}
$\\\hline
\end{tabular}
\\
\end{gather*}

Some points should be observed. First, as $SU(3)_{C}\otimes SU(2)_{L}\otimes
U(1)_{Y}$ is a maximal sub-group of $SU(5),$ then we have $SU(3)_{C}\otimes
SU(2)_{L}\otimes SU(2)_{R}\otimes U(1)_{Y}$ $\subset SU(5)\otimes
SU(2)_{R}\subset SU(7)$. In fact \cite{Slansky} $SU(5)\otimes SU(2)\otimes
U(1)_{X}$ \ is a maximal sub-group of \ $SU(7)$ and we can  assume $SU(2)$
to have the right chirality $SU(2)_{R}$.

A second point is that the mass terms of leptons $\overline{l_{eL}}\chi
_{L}e_{R}$ require Higgs representations $\chi_{L}\sim(\mathbf{1,2,1,}%
1\mathbf{\,).}$ Similarly the mass terms of the mirror partners $\overline
{L_{ER}}\chi_{R}E_{L},$ require $\chi_{R}\sim(\mathbf{1,1,2,}1\mathbf{\,).}$
Mixing terms of the type $\overline{e_{R}}S_{D}E_{L},$ $\overline{\nu_{R}%
}S_{D}N_{EL}$ need $S_{D}\sim$ $(\mathbf{1,1,1,}0\mathbf{\,).}$ Mass terms of
the Majorana type $\overline{l_{eL}}\widetilde{\chi_{L}}N_{EL}^{C}$ need
$\widetilde{\chi_{L}}\sim$ $(\mathbf{1,2,1,}-1\mathbf{\,)}$ and $\overline
{L_{ER}}\widetilde{\chi_{R}}\nu_{eR}^{C}$ \ need $\widetilde{\chi_{R}}\sim$
$(\mathbf{1,1,2,}-1\mathbf{\,)}$ in order to give mass to neutrinos. The
$\overline{N_{EL}^{C}}S_{M}N_{EL}$ and $\overline{\nu_{eR}^{C}}S_{M}\nu_{eR}$
terms are possible with $S_{M}\sim(\mathbf{1,1,1},0).$ Now, let us search for
the representations of $\chi_{L,R},$ $S_{D}$ and $S_{M}$ in the $SU(7)$
context \cite{SU(7)}. The fermionic multiplet  are in the anomaly free
combination\footnote{We are using $\{\qquad\}$ for the $SU(7)$ components
too.} \cite{K. Yamamoto} $\mathbf{\{1\}\oplus\{7\}\oplus\{21\}\oplus\{35\}}$
corresponding to the spinor representation $\mathbf{64}$\textbf{ }of $SO(14)$
into \ which $SU(7)$ is embedded. In the previous multiplets, $\left\{
\mathbf{21}\right\}  $ is a 2-fold, $\left\{  \mathbf{35}\right\}  $ is a
4-fold and $\left\{  \mathbf{7}\right\}  $ is a 6-fold of totally antisymmetric
tensors. \ 

Let us note that $\mathbf{64}$ can  contain two families of ordinary
fermions with its respective mirror partners, for example the electron and
muon families as is showed in  Table 3. The other families can be
incorporated into other $\mathbf{64\ }$spinorial representation. The branching
rules for each component of the spinorial representation, under its sub-group
$SU(5)\otimes SU(2)_{R}$ , are \cite{Jihn E. Kim}: \textbf{ }
\begin{gather}
\mathbf{\{35\}}=[\mathbf{10}^{\ast}\mathbf{,1}]\oplus\lbrack\mathbf{10,2}%
]\oplus\lbrack\mathbf{5,1}],\nonumber\\
\mathbf{\{21\}}=[\mathbf{10}^{\ast}\mathbf{,1}]\oplus\lbrack\mathbf{5}^{\ast
}\mathbf{,2}]\oplus\lbrack\mathbf{1,1}],\\
\mathbf{\{7\}}=[\mathbf{5,1}]\oplus\lbrack\mathbf{1,2}],\nonumber
\end{gather}
and under $SU(3)_{C}\otimes SU(2)_{L}\otimes SU(2)_{R}\otimes U(1)_{Y}$ are
\begin{gather}
\mathbf{\{35\}}=%
\genfrac{}{}{0pt}{0}{\underbrace{\mathbf{(1,1,1},\mathbf{-}2\mathbf{)}}%
}{e_{R}}%
\oplus%
\genfrac{}{}{0pt}{0}{\underbrace{\mathbf{(3,1,1},4/3\mathbf{)}}}{u_{R}}%
\oplus%
\genfrac{}{}{0pt}{0}{\underbrace{\mathbf{(3,2,1},1/3)}}{\dbinom{c}{s}_{L}}%
\mathbf{\oplus}%
\genfrac{}{}{0pt}{0}{\underbrace{\mathbf{(1,1,1},-2\mathbf{)}}}{E_{L}}%
\oplus\nonumber\\%
\genfrac{}{}{0pt}{0}{\underbrace{\mathbf{(1,1,1},-2\mathbf{)}}}{M_{L}}%
\oplus%
\genfrac{}{}{0pt}{0}{\underbrace{\mathbf{(3,1,1,}4/3)}}{U_{L}}%
\mathbf{\oplus}%
\genfrac{}{}{0pt}{0}{\underbrace{\mathbf{(3,1,1,}4/3)}}{C_{L}}%
\mathbf{\oplus}%
\genfrac{}{}{0pt}{0}{\underbrace{\mathbf{3,1,1},-2/3\mathbf{)}}}{s_{R}}%
\oplus\nonumber\\%
\genfrac{}{}{0pt}{0}{\underbrace{\mathbf{(1,2,1,-}1\mathbf{)}}}{\dbinom
{\nu_{e}}{e}_{L}}%
\oplus%
\genfrac{}{}{0pt}{0}{\underbrace{\mathbf{(3,1,2},1/3)}}{\dbinom{U}{D}_{R}}%
\mathbf{\oplus}%
\genfrac{}{}{0pt}{0}{\underbrace{\mathbf{\mathbf{(3,1,2},}%
1/3\mathbf{\mathbf{)}}}}{\dbinom{C}{S}_{R}}%
\mathbf{,}%
\end{gather}%
\begin{align}
\mathbf{\{21\}}  &  =%
\genfrac{}{}{0pt}{0}{\underbrace{\mathbf{(1,1,1},-2)}}{\mu_{R}}%
\mathbf{\oplus}%
\genfrac{}{}{0pt}{0}{\underbrace{\mathbf{(3,1,1},4/3)}}{c_{R}}%
\mathbf{\oplus}%
\genfrac{}{}{0pt}{0}{\underbrace{\mathbf{(3,2,1},1/3)}}{\dbinom{u}{d}_{L}}%
\mathbf{\oplus}%
\genfrac{}{}{0pt}{0}{\underbrace{\mathbf{(1,1,2},-1)}}{\dbinom{N_{E}}{E}_{R}}%
\mathbf{\oplus}\nonumber\\
&
\genfrac{}{}{0pt}{0}{\underbrace{\mathbf{(1,1,2},-1)}}{\dbinom{N_{M}}{M}_{R}}%
\mathbf{\oplus}%
\genfrac{}{}{0pt}{0}{\underbrace{\mathbf{(3,1,1},-2/3)}}{D_{L}}%
\mathbf{\oplus}%
\genfrac{}{}{0pt}{0}{\underbrace{\mathbf{(3,1,1},-2/3\mathbf{)}}}{S_{L}}%
\oplus%
\genfrac{}{}{0pt}{0}{\underbrace{(\mathbf{1,1,1},0)}}{N_{ML}}%
,
\end{align}

\begin{align}
\mathbf{\{7\}}  &  =%
\genfrac{}{}{0pt}{0}{\underbrace{\mathbf{(1,2,1},-1)}}{\dbinom{\nu_{\mu}}{\mu
}_{L}}%
\mathbf{\oplus}%
\genfrac{}{}{0pt}{0}{\underbrace{\mathbf{(3,1,1,}-2/3)}}{d_{R}}%
\mathbf{\oplus}%
\genfrac{}{}{0pt}{0}{\underbrace{\mathbf{(1,1,1},0\mathbf{)}}}{N_{EL}}%
\mathbf{\mathbf{\oplus}}%
\genfrac{}{}{0pt}{0}{\underbrace{\mathbf{\mathbf{(1,1,1},}0\mathbf{\mathbf{)}%
}}}{N_{ML}}%
\mathbf{.}\\
\{\mathbf{1\}}  &  \mathbf{=}%
\genfrac{}{}{0pt}{0}{\underbrace{(\mathbf{1,1,1,}0)}}{\mathbf{\ }\nu_{eR}}%
\mathbf{\ }.
\end{align}
From the product $\{\mathbf{63}\}\otimes\{\mathbf{63}\}=\{\mathbf{1}%
\}+\{\mathbf{63}\}\oplus...,\ $we obtain the Higgs representations producing
the mass terms for the fermions in the spinorial multiplet $\{\mathbf{63}%
\}=\mathbf{\{7\}\oplus\{21\}\oplus\{35\}}$ of $SU(7).$ With the help of the
branching rules (15) - (17), we take
\begin{gather}
\chi_{L}\sim\mathbf{\{7}^{\ast}\mathbf{\}\supset(1,2,1,}1\mathbf{),\quad}%
\chi_{R}\sim\mathbf{\{21}^{\ast}\mathbf{\}\supset(1,1,2,}1\mathbf{),}\\
S_{D}\sim\{\mathbf{21\}\supset\,}(\mathbf{1,1,1,}0),\quad S_{M}\sim
\{\mathbf{1\}\sim(1,1,1,}0\mathbf{).}%
\end{gather}
Finally we can have the following breaking chain with two singlets and two
doublets of Higgs representations:%

\begin{gather}
SU(7)\ \underrightarrow{S_{M}}\ SU(5)\otimes SU(2)_{R}\otimes\mathcal{D}%
\ \underrightarrow{S_{D}}\ G_{SM}\otimes SU(2)_{R}\nonumber\\
\ \underrightarrow{\chi_{R}}\ SU(3)_{C}\otimes SU(2)_{L}\otimes U(1)_{Y}%
\ \underrightarrow{\chi_{L}}\ SU(3)_{C}\otimes U(1)_{e.m}\ .
\end{gather}
The component of $\phi^{\alpha\beta}=\{\mathbf{21\}}$ that breaks $\mathcal{D}
$-parity is given by $S_{D}=\phi^{67}$ which is odd under $\mathcal{D}$-parity
\cite{Kim} and the $S_{M}$ field  being an $SU(7)$ singlet it preserves $\mathcal{D}$-parity. 

We can  write a n$SU(7)$ invariant Higgs potential similar to the one given in  Equation.(12)
with the obvious changes $\mathbf{144\rightarrow\{7}^{\ast}%
\mathbf{\},45\rightarrow}\{\mathbf{21\},}$ and $\mathbf{54\rightarrow
}\{\mathbf{1\}}$.
\section{Conclusions.}
In conclusion we have shown that parity can be spontaneously broken by a
simple Higgs sector with two doublets and two singlets. The grand unified
sector that contains this possibility is more restricted than other scenarios.
One of the new proposed singlets can have a breaking scale not very far from
the Fermi scale. A similar conclusion was recently found in a different
approach for two doublets models \cite{MAR}. In the case of the model with
mirror fermions it is possible a significant contribution to the magnetic moment of  
electrons and muons \cite{Chavez} due to  couplings with the mirror fermions of
the type $\mathcal{\,}f(\overline{l_{L}}\chi_{L}e_{R}+\overline{l_{R}}\chi
_{R}E_{L})+$ $f^{\,\,\prime}\,\overline{e_{R}}E_{L}S_{D}$. This terms can give an
important contribution to the muon anomaly and will be connected to the
breaking of the Weinberg symmetry \cite{Chavez et al}. 

\textit{Acknowledgments:} We thanks CNPq and FAPERJ for financial support.

\end{document}